\documentclass[a4paper]{article}

\usepackage{INTERSPEECH2022}
\usepackage{graphicx} 
\usepackage{graphbox}
\usepackage{caption,subcaption}
\usepackage{cite}
\usepackage{xspace}
\usepackage{tabularx}
\usepackage{booktabs}
\usepackage{multirow}
\usepackage{makecell}
\usepackage{hyperref}
\hypersetup{
    colorlinks=true,
    linkcolor=blue,
    filecolor=magenta,      
    urlcolor=cyan,
    citecolor=blue,
}
\usepackage{colortbl}
\definecolor{Dejan}{rgb}{1.0,0.0,0.0}
\definecolor{AlexD}{rgb}{0.5,0.0,1.0}
\definecolor{AlexR}{rgb}{0.0,0.0,0.8}

\usepackage{booktabs, multirow}
\usepackage{soul}
\usepackage[table]{xcolor}
\usepackage{changepage,threeparttable} 
\usepackage{pifont}
\usepackage{fontawesome}

\title{Implicit Neural Spatial Filtering for Multichannel Source Separation in the Waveform Domain}
\name{Dejan Markovi\'c$^1$, Alexandre D{\'e}fossez$^2$, Alexander Richard$^1$}

\address{
  $^1$Meta, Reality Labs Research, Pittsburgh PA, USA\\
  $^2$Meta AI Research, Paris, France}
\email{\{dejanmarkovic,defossez,richardalex\}@fb.com}

\begin{document}

\maketitle
\begin{abstract}
\vspace{-0.05in}
We present a single-stage casual waveform-to-waveform multichannel model that can separate 
moving sound sources based on their broad spatial locations in a dynamic acoustic scene. 
We divide the scene into two spatial regions containing, respectively, the target and the interfering sound sources. The model is trained end-to-end and performs spatial processing implicitly, without any components based on traditional processing or use of hand-crafted spatial features. We evaluate the proposed model on a real-world dataset and show that the model matches the performance of an oracle beamformer followed by a state-of-the-art single-channel enhancement network. 
\end{abstract} 
\vspace{-0.1in}

\section{Introduction}
\vspace{-0.05in}
Spatial clues captured by left and right ear help us localize sound sources and focus attention to directions of interest. Similarly, multiple microphones can be used to separate sound sources based on their spatial locations. With modern devices often featuring two to eight microphones, multichannel processing is becoming increasingly important in commercial applications. 
At the same time, the rise of virtual and augmented reality is giving rise to novel application scenarios, and introducing new challenges in using multiple microphones to separate target sounds from complex soundscapes.       

While deep neural networks (DNNs) represent the state-of-the-art in the single-channel source separation, denoising, and enhancement \cite{wang2018}, mainstream multichannel separation still relies on traditional array processing techniques \cite{gannot2017}. Spatial filtering, a.k.a. beamforming, exploits spatial sampling by multiple microphones in order to enhance signals coming from a given  location in space. An example of adaptive spatial filter is the widely used Minimum Variance Distortionless Response (MVDR) beamformer \cite{habets2010}, whose original formulation dates back to a half-century ago \cite{capon1969}.

More recently, DNNs have been applied to multichannel processing with a varying degree of integration with traditional beamformers. Proposed approaches employed neural networks alongside traditional filters \cite{chen2017, wu2020, fu2021,liu2022}, used DNNs to estimate parameters of traditional filters  \cite{nugraha2016, erdogan2016, heymann2016, zhang2017, erdogan2016b, xiao2017, heymann2017, qian2018, wang2018b,  ceolini2019, xu2020, zhang2020b, zhang2021, gu2021, chakrabarty2019, ochiai2017, xu2021}, or directly predicted filter coefficients \cite{xiao2016, ochiai2017, ren2021, sainath2017, luo2019, chakrabarty2019, luo2020, xu2021}. These spatial filters were often followed by post-filter enhancement networks \cite{zhang2017,wang2020,wang2021}. In the last couple of years, however, there has been increasing interest in using DNNs for spatial processing without explicitly using or generating spatial filters. Approaches have been proposed for both speech denoising \cite{tawara2019, li2019, tolooshams2020, liu2020, tzirakis2021, pandey2022} and separation \cite{gu2019, gu2020, zhang2020, jenrungrot2020}. For separation purposes, \cite{gu2019, gu2020, zhang2020} used hand-crafted or learned spatial features as inputs, whereas \cite{jenrungrot2020} performs beam steering as a separate step in an iterative localization-separation procedure. 

In this work we take a different approach to spatial source separation. We observe that often there is a  certain natural separation between regions containing target and interfering sources. In case of a video call, for example, the speaker is usually in front of the communication device. Similarly, in case of a VR collaboration platform, we are interested in capturing sounds generated by the user wearing the VR headset. We can also envision a scenario in which AR glasses are used to enhance sounds of people sitting at a table in a noisy environment.
Given this insight, in this work we divide the space into two predefined regions containing, respectively, target and interfering sources. 
The distinction between regions where to render soundfields and the regions where to minimize interference has been made for sound reproduction purposes \cite{betlehem2015}, but it has not been used for the mirror task of sound capture. 
By performing spatial filtering on regions instead of point-like locations in space, we simplify the treatment of dynamic scenes. 
Furthermore, we observe that in multistage processing frameworks, later stages often do not have access to all information from the input. For example, enhancement post-filter that follows a beamformer does not have access to spatial information that's lost after spatial filtering.  
Intuitively, a joint separation in spatial, spectral and temporal domain has potential to be more efficient. 
Therefore, unlike most previous works, we approach the task using a single stage approach and do it directly in the waveform domain. 
Our multichannel model is based on Demucs architecture that has shown state-of-the-art performance in both music source separation \cite{defossez2019, defossez2021} and speech enhancement \cite{defossez2020}, while working in real-time on a single CPU core.  
In order to teach the network to implicitly use spatial information, we devise the following training strategy: we train the network on dynamic scenes with moving sources of the same type; and since the only discriminative feature between target and interfering sources is their presence in one of the two regions, the network has to understand the spatial configuration in order to perform the separation task. 
By discriminating between the target and interference sound sources, the network is implicitly performing source localization.
Finally, we note that, although in this work we focus on speech signals and we use left/right, and near/far spatial subdivisions, the proposed approach can be extended to other types of sounds and space subdivisions. 
\begin{figure*}[t]
	\centering
	\includegraphics[width=1.8\columnwidth]{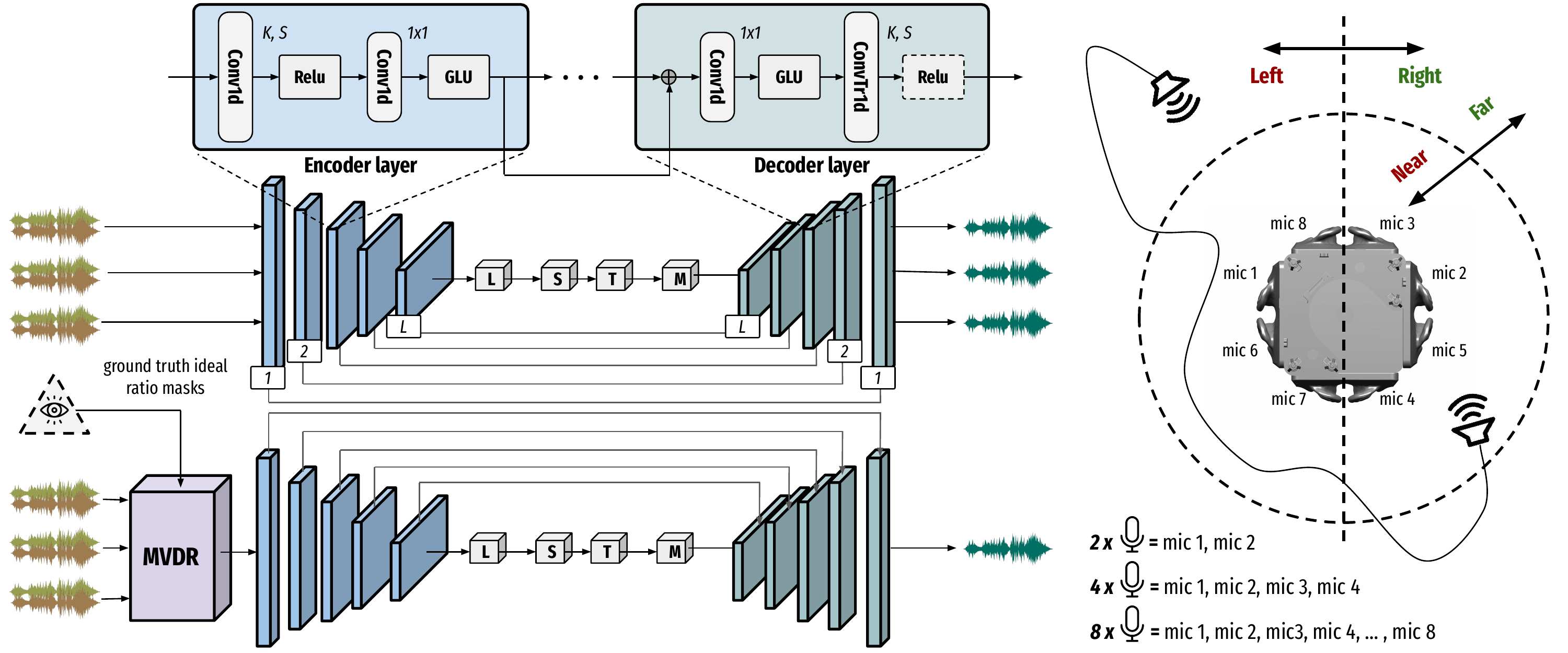} 
	\caption{\textbf{Top left}: multichannel Demucs architecture. \textbf{Bottom left}: MVDR beamformer followed by single-channel Demucs. \textbf{Right}: microphone array geometry and space subdivision.}
    \label{fig:demucs}
	\vspace{-0.4cm}
\end{figure*}

We compare the proposed approach against a mask-based MVDR oracle beamformer followed by a single-channel Demucs post-filter. 
Training and evaluation is performed on real-world data. 
Experimental results show that although it performs a difficult task of separating real-world moving sources while working directly on raw waveforms, 
the proposed approach is comparable to the state-of-the-art oracle system. We analyze model performance for different number of microphones and sources, and we show that the model is indeed performing separation in space other than time and frequency alone.     
\vspace{-0.05in}

\section{Related work}
\vspace{-0.05in}
Most works on neural multichannel processing incorporate networks into designs of traditional beamforming filters.
In \cite{nugraha2016} DNNs were used to model source spectra needed for the multichannel Wiener filter. Similarly, networks were used to estimate spectral masks and compute spatial covariances needed for generalized eigenvalue (GEV) \cite{erdogan2016b, heymann2016, heymann2017, xu2021} or MVDR \cite{erdogan2016b, heymann2016, erdogan2016, zhang2017, xiao2017, ochiai2017, wang2018b, ceolini2019, chakrabarty2019, xu2020, zhang2021, gu2021} beamformers. 
Other works combined neural networks with a set of predefined fixed beamformers steered towards a discreet number of directions \cite{chen2017, wu2020, fu2021,liu2022}.
Around the same time, networks that directly estimate spatial filters were developed in both frequency \cite{xiao2016, ochiai2017, xu2021, ren2021} and time \cite{sainath2017, luo2019} domain. 
It is worth noting that many of these works were developed for automatic speech recognition (ASR) task, integrating acoustic models with proposed beamforming networks \cite{erdogan2016b, xiao2016, xiao2017, ochiai2017, heymann2017, sainath2017, zhang2020b}. Unlike communication systems, ASR systems do not require low-latency, causal processing.

Recently, the first steps towards end-to-end implicit neural beamforming have been undertaken. 
Several works used time-frequency representation of multichannel inputs in order to estimate real \cite{li2019} or complex \cite{tolooshams2020, gu2021, tzirakis2021} spectral ratio masks for reference channels. 
In \cite{wang2020,wang2021} DNNs are used to directly predict the real and imaginary parts of target spectograms but they still employ traditional MVDR beamformers and DNN post-filters in a multistage processing framework.
On the other hand, in \cite{pandey2022} a two-stage approach was shown to perform well without MVDR, though it focused on speech enhancement of static sources. 
Approaches working in time domain have been developed as well. 
Speech denoising was performed using a time-domain convolutional autoencoder \cite{tawara2019}, or employing a residual architecture with a fully convolutional network \cite{liu2020}.
The single-channel Conv-TasNet \cite{luo2019b} has been extended to incorporate spatial information by using hand-crafted \cite{gu2019} or learned \cite{gu2020, zhang2020} spatial features obtained from the multichannel data. Similarly to our work, Demucs was used for multichannel source separation in \cite{jenrungrot2020}, but instead of implicitly using the spatial information, steering of beams towards desired directions was performed before the channels were fed to the network. 
\vspace{-0.1in}

\section{Approach}
\vspace{-0.05in}
Extracting a target sound source from a mixture usually involves some sort of visual or acoustic source localization or estimation of the acoustic transfer functions, which are then used to steer the beams, i.e., select or build spatial filters for the given locations. In dynamic scenes these estimates need to be constantly updated and, as a consequence, the performance degrades significantly. Given difficulties in dealing with dynamic scenes, most current research into implicit neural spatial processing focuses on static scenes. In this work we specifically consider moving sound sources. In particular, instead of focusing on point-like positions in space, we divide the space into two predefined regions containing, respectively, target and interfering sources, with no constraints on source movement inside these regions (i.e., regions are fixed but the sources are free to move). The model is trained to preserve sounds coming from the target region while suppressing sources in the interference region. Below we discuss the model architecture, training strategy and describe the dataset used to train the proposed approach. 
\vspace{-0.1in}

\subsection{Model architecture}
\vspace{-0.05in}
Our model is based on the causal Demucs architecture presented in \cite{defossez2020}, a multi-layer convolutional encoder-decoder with U-Net skip connections and a LSTM for long range dependencies, that works directly in the waveform domain. Originally proposed for music source separation \cite{defossez2019}, Demucs showed state-of-the-art performance on a single-channel speech enhancement task while working in real time on a single CPU core \cite{defossez2020}. 
This makes Demucs architecture particularly interesting for the task of spatial source separation. The architure was mostly unchanged, except for the number of input/output channels and the fact that we did not use the resampling used in~\cite{defossez2020}. For completeness, we give a brief description hereafter.

The Demucs architecture is shown at left of Fig.~\ref{fig:demucs}. It is characterized by its number of layers $L$, initial number of hidden channels $H$, layer kernel size $K$ and stride $S$. The encoder and decoder layers are numbered from 1 to $L$ (in reverse order for the decoder, so layers at the same scale have the same index). The $i$-th encoder layer consists in a convolution with kernel size $K$ and stride $S$, and $2^{i} H$ output channels. We use a ReLU activation, followed by
a second ``1x1'' convolution, i.e. with a kernel size of 1, stride of 1, and $2^{i+1}H$ channels as output. A GLU
activation function is used to take back the number of channels to $2^{i} H$.
After the encoder, a LSTM is applied with $2^{L-1} H$ hidden channels and 2 layers.
The decoder is built symmetrically, receiving as input the sum of the output of the matching encoder layer and of the previous decoder layer or the LSTM for the first one.

We used $L{=}5$, $H{=}64$, $K{=}8$ and $S{=}4$. The model is fed
with audio sampled at 48kHz and has $C$ input/output channels, which can be 1 when using beamforming, or set to the number of microphones in the array. With those parameters, the Demucs architecture is causal, with an initial lag of 74ms. Unlike most other approaches that use multi-channel data to produce a single-channel output, we ask the network to output target signals at each channel. By doing so, we preserve the spatial information of the target sources for subsequent downstream tasks.
\vspace{-0.2in}

\subsection{Training}
\vspace{-0.05in}
The proposed model does not have any component based on traditional spatial processing and does not use any hand-crafted spatial features. The architecture itself was successfully employed for speech denoising, enhancement and source separation. As such, there is a possibility that the model will not learn to exploit spatial information at all. In order to enforce extraction and use of 
spatial information from the multichannel input, we train the network to perform a task that requires spatial understanding of the acoustic scene. In particular, during training the network is presented with two sound sources of the same type and asked to extract one of them and suppress the other. The sources are freely moving in the environment and the only discriminative feature between the two is their presence in the respective spatial regions. By discriminating between target and interference, the network is implicitly performing the source localization. 
Finally, we train the model by minimizing the L1 loss on the raw waveforms for all microphone channels.

\vspace{-0.1in}
\subsection{Dataset}
\vspace{-0.05in}
We use an internal, real-world dataset to train and evaluate our model. In order to capture enough spatial diversity we use data captured by multiple 3Dio Omni binaural microphone arrays, each made by 8 microphones forming 4 pairs of binaural ears (see right of Fig.~\ref{fig:demucs}). Note that in this work we focus on generic arrays; we used binaural microphone rigs simply due to data availability reasons. We test our system considering the rigs as either a 2, 4, or 8-unit microphone array as illustrated in Fig.~\ref{fig:demucs}. Captured data consists of a single loudspeaker carried by a person moving freely in the environment. The loudspeaker plays speech signals from the VCTK corpus \cite{vctk}. The positions of the microphone arrays and the loudspeaker are known and actively tracked at 240 frames per second using the Optitrack system. The loudspeaker movement covers a distance of 4.6 m horizontally and 2.4 m vertically. We recorded 42 hours of audio data with 27 rigs distributed in the environment. The audio was captured at 96kHz and then downsampled to 48kHz.

After capture, the multichannel recordings are divided into segments of 3 seconds and each of them is classified as either target or interference depending on the loudspeaker position with respect to the given array (we discard ambiguous segments in which the speaker crosses from one region to the other). In this work we consider two space subdivisions: left-right split and near-far split with 0.7 m being the boundary between near and far field. 
During training, the input mixture is obtained by pairing the target recording with a randomly selected interference recording. The gain of the two recordings is adjusted randomly and the signals are combined to form the input mixture. The model is trained to minimize L1 loss between the target recording and network output given this mixture as input. 
\vspace{-0.05in}

\section{Experiments}
\vspace{-0.05in}
\subsection{Setup}
\vspace{-0.05in}
\noindent{\textbf{Competing systems:}}
We compare the proposed approach against an oracle MVDR beamformer and a single-channel Demucs trained on the same dataset using the output of the oracle MVDR as a the input signal, as shown at bottom left of Fig.~\ref{fig:demucs}. Oracle MVDR has access to ground truth ideal ratio masks used to compute spatial covariances. 

\noindent{\textbf{Evaluation metrics:}}
We compute the scale invariant signal-to-distortion ratio (SI-SDR),  
and the Mel $ \ell_2 $ loss, i.e., the $ \ell_2 $ distance of ground truth and predicted signal in Mel spectral domain.
For fair comparison with a single channel model we compute the metrics on the first microphone only. MVDR uses the first microphone as a reference channel. 

\noindent{\textbf{Spatial subdivision:}} Two space subdivisions are considered: (1) Two half-spaces -- target sources are on the right and interfering sources on the left; (2) Near/far split -- target sources are in far-field and interfering sources in near-field of the array, with 0.7 m as the boundary (note that we choose far sources as targets since this is the more challenging scenario).   
\vspace{-0.05in}

\begin{table}[!htp]
    \centering
    \caption{Performance of different models with 2, 4, or 8 microphones. For (a), the input mixture has -0.9dB SI-SDR, for (b) it has -13dB SI-SDR. Under ``out'' we denote the SI-SDR of the output, with ``out-in'' we denote the SI-SDR improvement over the input. Mel loss is the $ \ell_2 $ distance of in Mel spectral domain.}
    \label{tab:main_results}
    \scriptsize
    \begin{tabularx}{0.47\textwidth}{cXrrr}
        \toprule
        \multirow{2}{*}{\# \faMicrophone}    & \multirow{2}{*}{Method}    & \multicolumn{2}{r}{SI-SDR [dB] $\uparrow$}       & \multirow{2}{*}{Mel $\ell_2$ $\downarrow$}    \\
                                             &                            & out               & out-in            &             \\
        \toprule
        \multicolumn{5}{c}{\textit{(a) target signal: right; interference: left}} \\
        \midrule
        \multirow{3}{*}{2}  & oracle MVDR               & $ 1.6 $           & $ 2.4 $           & $ 0.955 $    \\
                            & oracle MVDR + 1ch~Demucs  & $ 2.5 $           & $ 3.3 $           & $ 0.860 $    \\
                            & Spatial Demucs            & $ 3.8 $           & $ 4.7 $           & $ 0.662 $    \\
        \midrule
        \multirow{3}{*}{4}  & oracle MVDR               & $ 3.5 $           & $ 4.4 $           & $ 0.930 $    \\
                            & oracle MVDR + 1ch~Demucs  & $ 4.2 $           & $ 5.0 $           & $ 0.788 $    \\
                            & Spatial Demucs            & $ 5.2 $           & $ 6.1 $           & $ 0.636 $    \\
        \midrule
        \multirow{3}{*}{8}  & oracle MVDR               & $ 5.5 $           & $ 6.4 $           & $ 0.911 $    \\
                            & oracle MVDR + 1ch~Demucs  & $ 5.8 $           & $ 6.7 $           & $ 0.740 $    \\
                            & Spatial Demucs            & $ 7.7 $           & $ 8.6 $           & $ 0.553 $    \\
        \toprule
        \multicolumn{5}{c}{\textit{(b) target signal: far; interference: near}} \\
        \midrule
        \multirow{3}{*}{2}  & oracle MVDR               & $ -7.2 $          & $  5.7 $          & $ 1.231 $    \\
                            & oracle MVDR + 1ch~Demucs  & $ -0.5 $          & $ 12.4 $          & $ 0.844 $    \\
                            & Spatial Demucs            & $  0.0 $          & $ 13.0 $          & $ 0.782 $    \\
        \midrule
        \multirow{3}{*}{4}  & oracle MVDR               & $ -2.1 $          & $ 10.8 $          & $ 1.111 $    \\
                            & oracle MVDR + 1ch~Demucs  & $  1.6 $          & $ 14.5 $          & $ 0.852 $    \\
                            & Spatial Demucs            & $  0.0 $          & $ 13.0 $          & $ 0.810 $    \\
        \midrule
        \multirow{3}{*}{8}  & oracle MVDR               & $  2.4 $          & $ 15.4 $          & $ 1.039 $    \\
                            & oracle MVDR + 1ch~Demucs  & $  4.5 $          & $ 17.5 $          & $ 0.809 $    \\
                            & Spatial Demucs            & $  1.1 $          & $ 14.1 $          & $ 0.693 $    \\
        \bottomrule
    \end{tabularx}
\end{table}

\vspace{-0.15in}
\subsection{Results}
\vspace{-0.05in}
\noindent{\textbf{Comparison with the MVDR oracles:}}
On Table~\ref{tab:main_results}, we compare the oracle MVDR with and without single-channel Demucs as post-processing, with proposed Spatial Demucs, the former having access to ground truth ideal ratio masks used for computation of spatial covariances, and the latter receiving no ground truth spatial information. As expected, adding single-channel Demucs as post-processing always improves the results of the beamforming only method. 
On the left-right case, using Spatial Demucs, brings a gain of 1.4 dB over the MVDR oracle with single channel denoising when using 2 mics arrays, and 2.1 dB for a 8 mics array.
When considering the far-near case, Spatial Demucs is competitive with the MVDR oracle with Demucs for a 2 mics array, although it lags behind by 3.4 dB when considering 8 mics array. We conjecture that the far-near case is more complex to separate when receiving no ground truth information, which gives a stronger edge to the beamforming based model, especially with a large number of microphones.
Finally, Spatial Demucs outperforms competing systems for all evaluated configurations in terms of Mel $ \ell_2 $ loss.


\begin{figure}[ht]
	\centering
	\includegraphics[width=0.8\columnwidth]{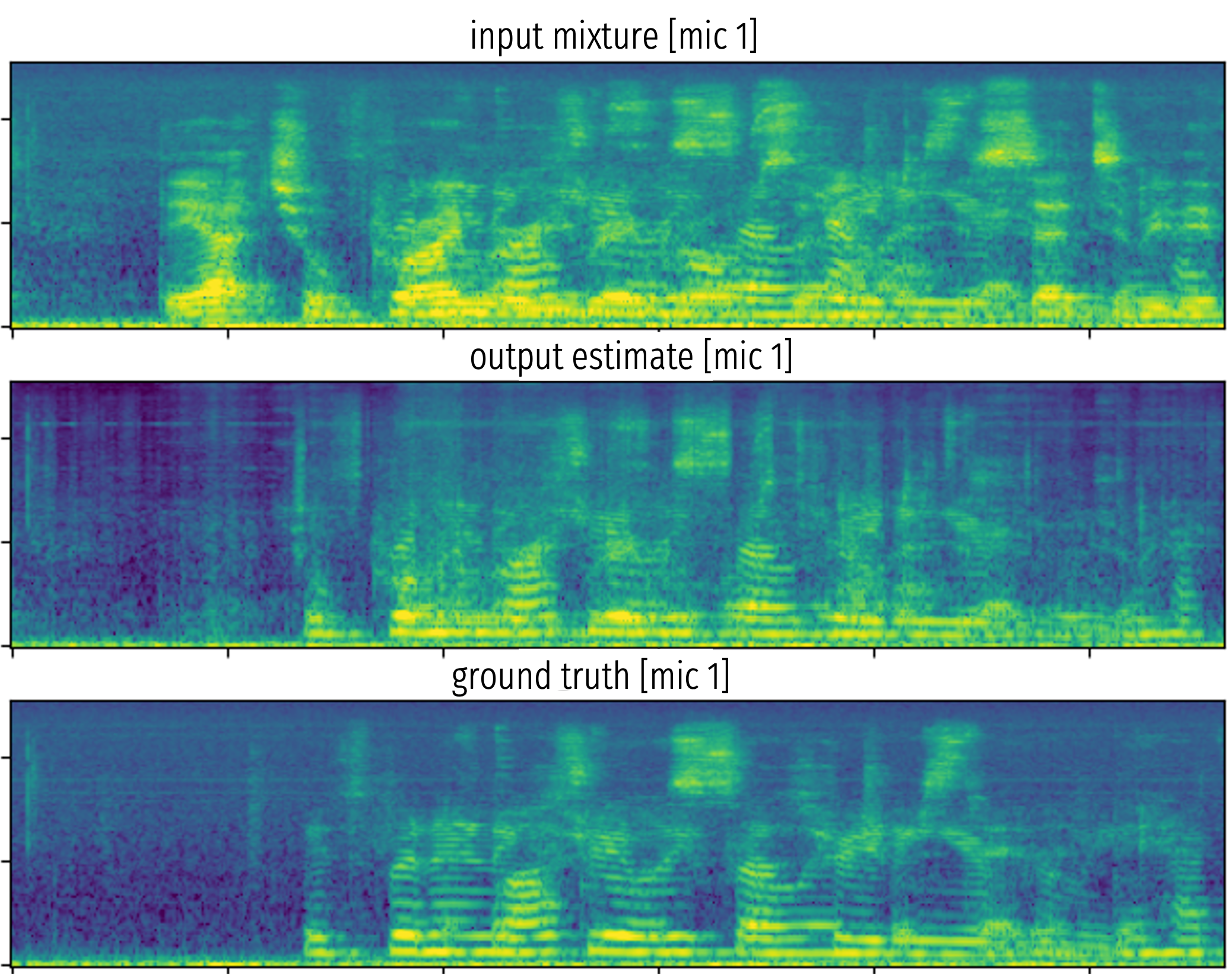} 
	\caption{Separation example with two interfering sources in the near-field and two target sources in the far-field. \label{fig:example}}
\end{figure}

\noindent{\textbf{Regions with multiple sounds sources:}}
We trained our model with one source per region. However, given that the model is asked to enhance / suppress sounds coming from the regions and not point-like positions in space, we expect it to generalize to multiple sources. Results for different number of sources in the two regions are shown in Table~\ref{tab:multi_source}. As expected, the model is able to perform sound separation even in these more challenging scenarios. An example of input, output and ground truth spectrograms is shown in Fig.~\ref{fig:example}. We observe that the model does a good job on extracting the desired sounds even with a high time-frequency overlap between sources. 

\begin{table}[!htp]
    \centering
    \caption{Performance of spatial demucs for different number of sources in the two regions. Results are reported for array with 2 microphones. Although trained on single sources, the system generalizes to multiple target and interference signals.}
    \label{tab:multi_source}
    \scriptsize
    \begin{tabularx}{0.47\textwidth}{ccXrrrrr}
        \toprule
        \multicolumn{2}{c}{\# \faVolumeUp \ \ per region}          & & \multicolumn{3}{c}{SI-SDR [dB] $\uparrow$}     & Mel $\ell_2$  $\downarrow$      \\
        \cmidrule(lr){1-2}                                  \cmidrule(lr){4-6}
        target       & interference                     & & in        & out       & out-in      & &                     \\
        \toprule
        \multicolumn{8}{c}{\textit{(a) target signal: right; interference: left}} \\
        \midrule
        1            & 2                                & & $ -4.9 $  & $ 0.6 $   & $ 5.5 $     & $ 0.587 $            \\
        2            & 2                                & & $ -1.1 $  & $ 3.3 $   & $ 4.4 $     & $ 0.491 $            \\
        2            & 3                                & & $ -3.1 $  & $ 1.5 $   & $ 4.7 $     & $ 0.519 $            \\
        3            & 3                                & & $ -1.2 $  & $ 2.8 $   & $ 4.0 $     & $ 0.470 $            \\
        \toprule
        \multicolumn{8}{c}{\textit{(b) target signal: far; interference: near}} \\
        \midrule
        1            & 2                                & & $ -17.8 $ & $  -9.0 $ & $ 8.8 $     & $ 0.984 $            \\
        2            & 2                                & & $ -13.6 $ & $  -5.5 $ & $ 8.1 $     & $ 0.845 $            \\
        2            & 3                                & & $ -16.1 $ & $ -10.2 $ & $ 5.9 $     & $ 1.066 $            \\
        3            & 3                                & & $ -13.7 $ & $  -8.0 $ & $ 5.7 $     & $ 0.931 $            \\
        \bottomrule
    \end{tabularx}
\end{table}

\begin{figure}[ht]
	\centering
	\includegraphics[width=0.8\columnwidth]{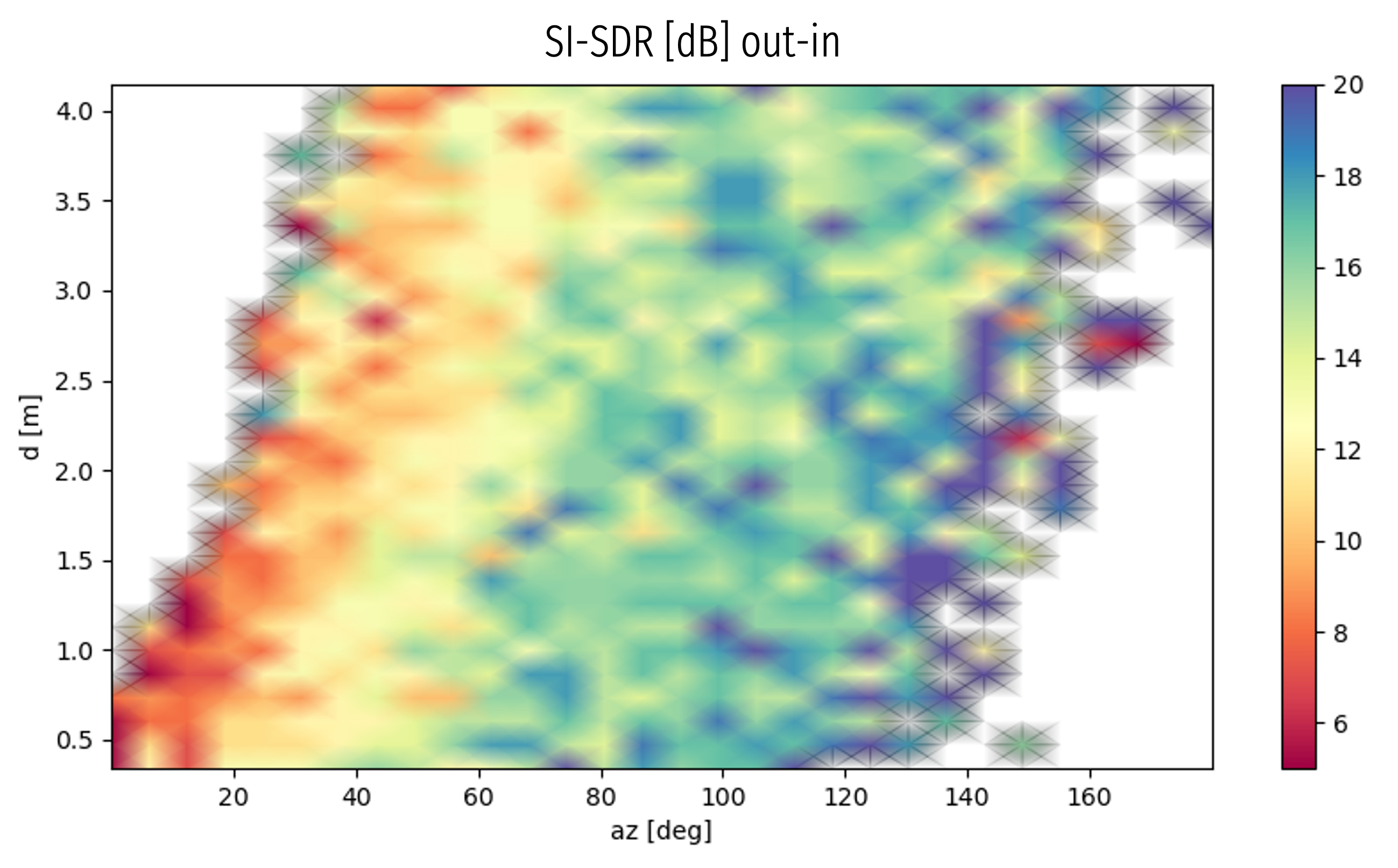} 
	\caption{SI-SDR as a function of angular and Cartesian distance between the sources. Results are reported for array with 8 microphones and far-near split. Dependence on spatial configuration indicates that the model exploits spatial information. \label{fig:spatial_analysis}}
	\vspace{-5mm}
\end{figure}

\noindent{\textbf{Spatial performance analysis:}}
In order to show that the proposed approach indeed uses spatial information for the source separation, we analyze the model performance for different spatial configurations between target and interfering sources. Fig.~\ref{fig:spatial_analysis} shows the variation of SI-SDR depending on the distance and the azimuth angle between the two sources. The worst performance is obtained when both the distance and angle between the two sources is small. Distance increase improves the performance except for small angles when interference occludes the target source. Intuitively, increasing the angular distance between the sources greatly improves the perforce. 

\noindent{\textbf{Training with synthetic data:}}
High amounts of real-world data may not be always available. To address this issue we show that a model trained on synthetic data can be fine tuned on small amounts of real-world data. Synthetic dataset is obtained by sampling a movement from the real dataset, and re-generating artificially the audio with simulated acoustic propagation and reverb. We train our model on synthetic data and fine-tune on 1 and 3 hours of data from real-world training set. We evaluate the performance on the real-world test set. The results in Table~\ref{tab:finetuning} show that fine tuning is effective. 

\begin{table}[!htp]
    \centering
    \caption{Performance of the model trained on synthetic data and fine tuned on 0, 1, and 3 hours of real data. For (a), the input mixture has -0.9dB SI-SDR, for (b) it has -13dB SI-SDR. Results are reported for array with 4 microphones.}
    \label{tab:finetuning}
    \scriptsize
    \begin{tabularx}{0.4\textwidth}{Xrrr}
        \toprule
        \multirow{2}{*}{real data}     & \multicolumn{2}{r}{SI-SDR [dB] $\uparrow$}       & \multirow{2}{*}{Mel $\ell_2$ $\downarrow$}    \\
                                                                         & out               & out-in            &             \\
        \toprule
        \multicolumn{4}{c}{   \textit{(a) target signal: right; interference: left   }} \\
        \midrule
        0 hrs                & $ -4.5 $           & $ -3.7 $          & $ 0.848 $    \\
        1 hrs                & $  5.0 $           & $ 5.8 $           & $ 0.722 $    \\
        3 hrs                & $  5.7 $           & $ 6.5 $           & $ 0.655 $    \\
        \toprule
        \multicolumn{4}{c}{\textit{(b) target signal: far; interference: near}} \\
        \midrule
        0 hrs               & $ -8.7 $          & $ 4.3 $          & $ 1.869 $    \\
        1 hrs               & $ -1.7 $          & $ 11.2 $         & $ 0.830 $    \\
        3 hrs               & $ -1.0 $          & $ 12.0 $         & $ 0.827 $    \\
        \bottomrule
    \end{tabularx}
\end{table}

\vspace{-0.05in}
\section{Conclusions}
\vspace{-0.05in}
We presented a method for training a multichannel neural model to perform spatial source separation. 
The approach divides the space into two regions containing, respectively, target and interfering sources. This spatial subdivision is fixed for a given network, however, we can envision using global conditioning in order to enable switching between several predefined configurations.
Experimental results show that the proposed approach achieves performance that is on par with a system that combines optimal beamformer with a state-of-the-art single-channel model. 
We show that the network is effectively exploiting spatial information present in the multichannel data.
Finally, although in this work we focus on speech signals, the extension to other source types is straightforward. 
\bibliographystyle{IEEEtran}

\bibliography{references}

\end{document}